\documentclass[twocolumn,amsmath,amssymb, nobibnotes, aps, prl, longbibliography,superscriptaddress]{revtex4-2}
\usepackage{graphicx}
\usepackage{lipsum}
\usepackage{xcolor}
\usepackage{bm}

\newcommand{\blue}{}

\renewcommand{\vec}[1]{\mathbf{#1}}
\newcommand{\unit}[1]{\hat{\mathbf{e}}_#1}

\begin{document}
\title{Topological Boundary Constraints in Artificial Colloidal Ice}
\author{Carolina Rodríguez-Gallo}
\affiliation{Departament de F\'{i}sica de la Mat\`{e}ria Condensada, Universitat de Barcelona, 08028 Spain}
\affiliation{Universitat de Barcelona Institute of Complex Systems (UBICS), Universitat de Barcelona, 08028 Spain}
\author{Antonio Ortiz-Ambriz}
\email{aortiza@fmc.ub.edu}
\affiliation{Departament de F\'{i}sica de la Mat\`{e}ria Condensada, Universitat de Barcelona, 08028 Spain}
\affiliation{Institut de Nanoci\`{e}ncia i Nanotecnologia, Universitat de Barcelona, 08028 Spain}
\author{Pietro Tierno}
\affiliation{Departament de F\'{i}sica de la Mat\`{e}ria Condensada, Universitat de Barcelona, 08028 Spain}
\affiliation{Universitat de Barcelona Institute of Complex Systems (UBICS), Universitat de Barcelona, 08028 Spain}
\affiliation{Institut de Nanoci\`{e}ncia i Nanotecnologia, Universitat de Barcelona, 08028 Spain}
\begin{abstract}
The effect of boundaries and how these can be used to influence the bulk behaviour in geometrically frustrated systems are both long-standing puzzles, often relegated to secondary role. 
Here we use numerical simulations and "proof of concept" experiments to demonstrate that boundaries can be engineered to control the bulk behavior in a colloidal artificial ice. 
We show that an antiferromagnetic frontier forces the system to rapidly reach the ground state (GS), as opposed to the commonly implemented open or periodic boundary conditions.
We also show that strategically placing defects at the corners generates novel bistable states, or topological strings which result from competing GS regions in the bulk. Our results could be generalized to other frustrated micro and nanostructures where boundary conditions may be engineered with lithographic techniques.  
\end{abstract}

\maketitle
In the thermodynamic limit the bulk properties of a statistical ensemble are no longer influenced by its boundaries. 
However, in frustrated spin systems the boundaries can induce configurations that propagate far into the bulk~\cite{han_phasespace_2009,han_phasespace_2010}.
{\blue Among several examples of frustrated systems in nature, the most representative one is spin ice~\cite{Harris1997,Ramirez1999,Bramwell2001}, which can be considered the magnetic "analogue" of the water ice~\cite{Giauque1936}.
Artificial spin ice systems (ASIs) based on lithographic engineering recently emerged as a versatile experimental platform to investigate geometric frustration effects~\cite{wang_artificial_2006a}. 
An ASI is composed by a lattice of nanoscale ferromagnetic islands, arranged to induce frustration~\cite{nisoli_colloquium_2013,cugliandolo_artificial_2017,
Skjarvo2020}.
In contrast to natural magnets, ASI allows to directly visualize the spins arrangement, 
a feature that has been used to investigate the effect of disorder~\cite{Branford2012,Chern2014,Drisko2017}, thermalization~\cite{Morgan2011,Farhan2013}
and degeneracy in many geometries~\cite{Wang2016,Gilbert2016,Canals2016,Perrin2016,Shi2018}.}
Alternative realizations include arrays of nanowires \cite{mengotti_realspace_2011}, 
patterned superconductors \cite{libal_creating_2009, latimer_realization_2013}, 
macroscopic magnets \cite{mellado_macroscopic_2012},
Skyrmions in liquid crystals \cite{ma_emergent_2016, duzgun_artificial_2019}, 
superconducting qbits \cite{king_quantum_2020},
and colloidal particles in bistable potentials \cite{ortiz-ambriz_engineering_2016}.

{\blue In such systems the presence of disorder or a finite temperature 
often prevents them from reaching the ground state (GS), and instead, they fall to a metastable state containing defects in form of charged vertices. 
These defects can be characterized by 
a topological charge $Q$ and have a topological nature, since they can only be destroyed when 
annihilating with other defects of opposite charge. 
In the GS, the vertices satisfy the 
ice rule, that prescribes a minimization of the local charge, $Q=0$.
While much attention has been placed on how temperature 
or external fields drive a system toward its GS, 
the role of boundaries in finite systems has been often overlooked.
This is of especial importance when dealing with interacting magnetic system where the interaction energies are governed by long-range dipolar forces.}

\begin{figure}[t]
\includegraphics[width=\columnwidth]{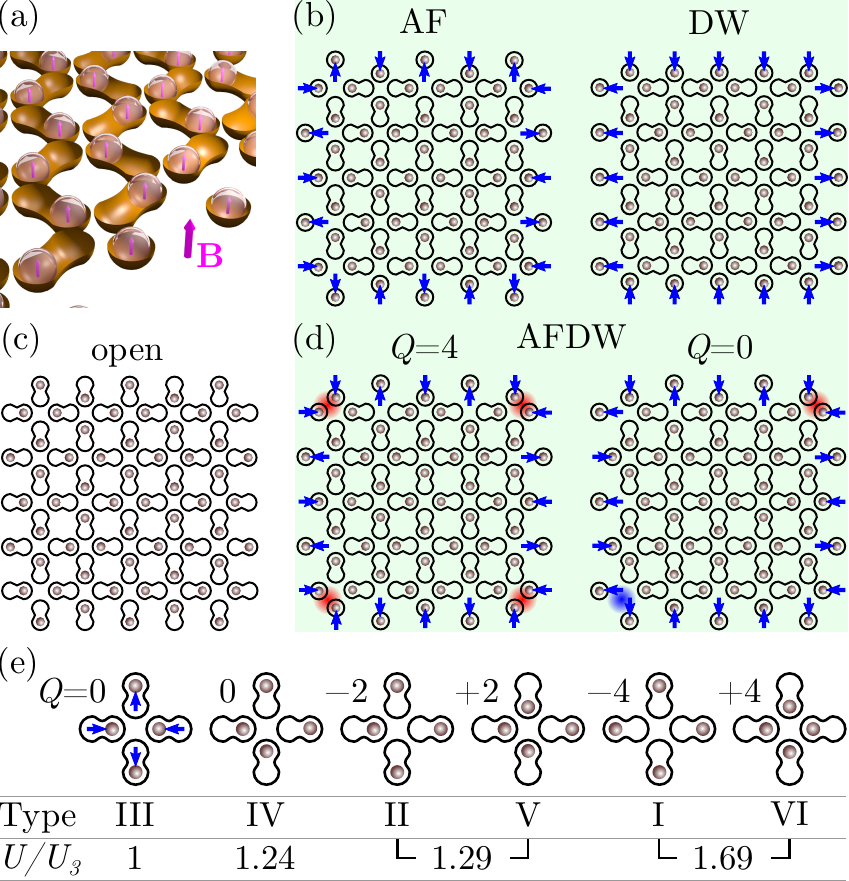}
\caption{(a-e) Different schematics showing: (a) 
The double-well geometry with paramagnetic colloids and the method of fixing the boundaries by using a single well. (b,c) 
The three basic types of boundary conditions, Antiferromagnetic (AF), Domain Wall (DW), and (c) open boundaries. (d) AFDW boundaries designed with defects which can carry a net charge ($Q=4$, left) or can be neutral ($Q=0$, right). The shaded region indicates boundaries which topologically protect the enclosed charge. (e) Different vertex types with their effective normalized energetic weight (bottom) and topological charge  (side). Vertices are ordered by increasing energetic ratio respect to Type III. The associated spins are shown on the Type III vertex.}
\label{fig:1}
\end{figure}

Here we show how boundaries can be engineered to control the bulk behaviour and the formation of topological states 
such as point defects and topological domain walls spanning the bulk. 
We demonstrate this concept here with an artificial colloidal ice, a system that recently emerged as a microscale soft-matter analogue to ASI~\cite{ortiz-ambriz_colloquium_2019}. 
Colloidal ice consists of an ensemble of 
paramagnetic colloids two dimensionally ($2$D) confined by gravity 
in topographic double wells, where the particles may sit in two 
stable positions and an external magnetic field $\bm{B}$
induces repulsive dipolar interactions, Fig.~\ref{fig:1}(a). 
One can assign a vector (analogous to a spin) to
each well such that it points
towards the vertex's centre (spin $in$) or away from it (spin
$out$), see Fig.~\ref{fig:1}.
When arranged in a square lattice, 
one can classify six possible vertex types
each of them with an associated topological charge $Q=2n-c_{N}$, where $n$ is the number of particles $in$
and $c_{N}$ the lattice coordination number. For the square is $c_{N}=4$.
Thus, vertices of type III and type IV have $Q=0$ and fulfil the ice rule, and type III gives rise to the GS. 
Topological defects are charged vertices with $Q\neq 0$, or closed loops of type IV vertices, Fig.~\ref{fig:1}(e).

\begin{figure*}[tp]
\includegraphics[width=\textwidth]{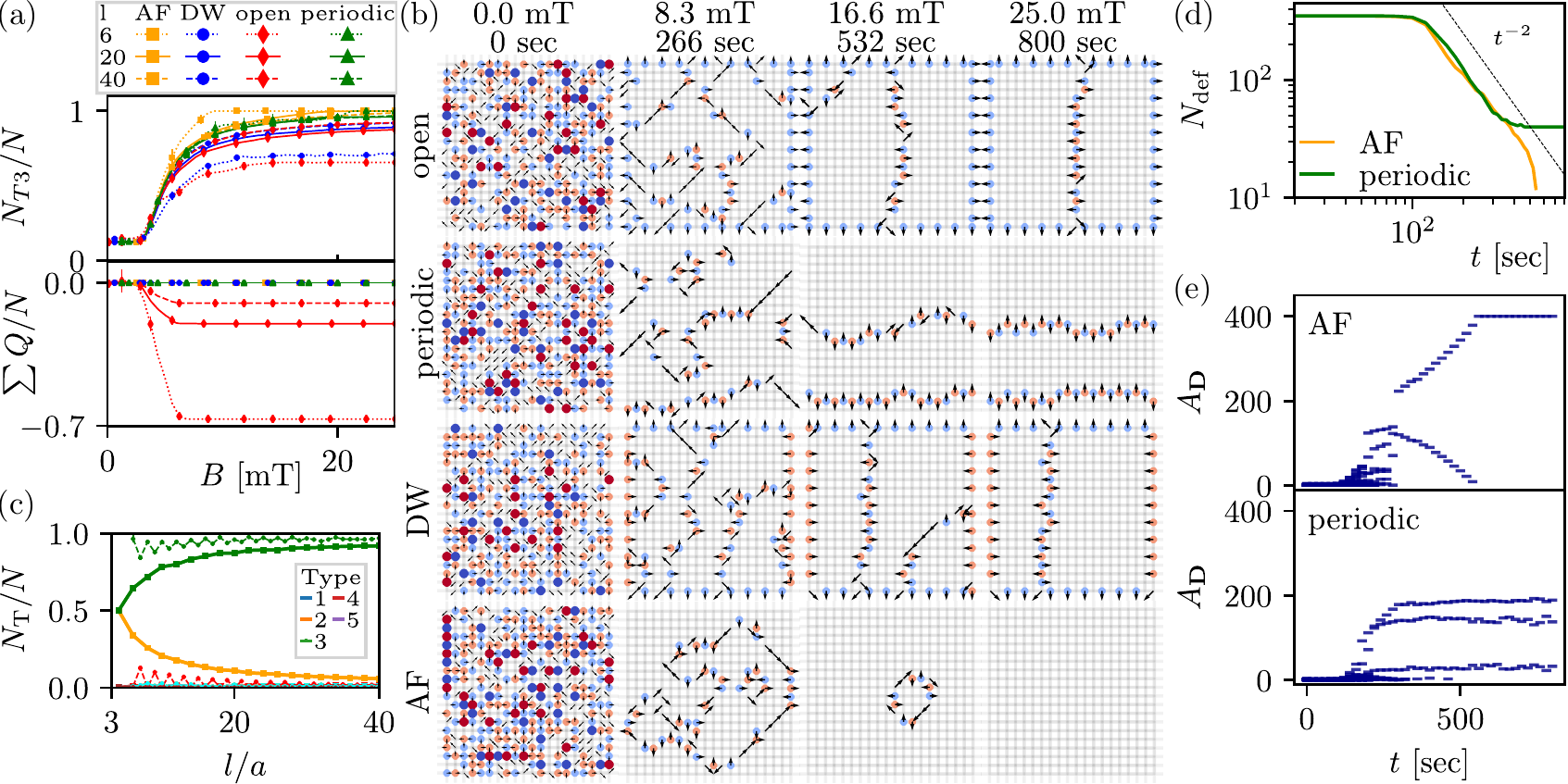}
\caption{(a) Fraction of type III vertices (Top)
and average vertex charge (Bottom) for different system sizes (numbers) and type of boundaries (symbols).  (b) Maps of defects for different types of charge-neutral boundaries. The colored circles indicate charges while colorless arrows indicate Type IV defects. The system width is $L = 20$ vertices.
(c) Fraction of vertices at the maximum field ($B=25$mT) versus system size, for open (solid lines) and periodic (dotted lines) boundaries.
{\blue (d) Number of defects $N_{\mathrm{def}}$ {\textit vs} time for the AF and periodic case, line denotes $t^{-2}$scaling. (e) Type III domain area $A_{\mathrm{D}}$ {\textit vs} time for AF (top) and periodic (bottom).}
}
\label{fig:2}
\end{figure*}

\begin{figure}[t!]
\includegraphics[width = \columnwidth]{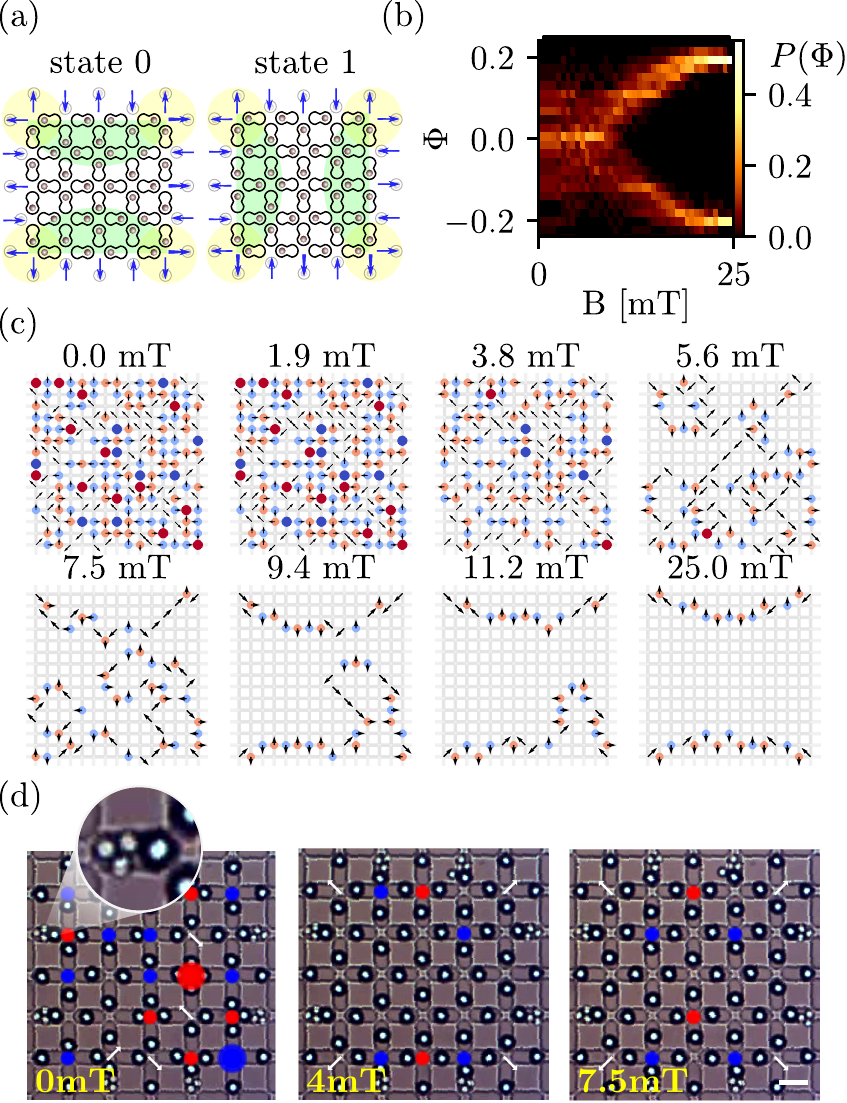}
\caption{(a) Schematics showing two states that appear after application of the field when using the AFDW configuration. Here topological charges can be connected either horizontally (right) or vertically (left). (b) Bifurcation of the order parameter observed for $N=100$ simulations. (c) Colormap showing the field induced symmetry breaking where the system chooses the state $0$. 
(d) Experimental observation of the first type of defect pattern, {\blue the second one is in the Supplemental Material~\cite{EPAPS}}. The enlargement shows the trap jammed by silica particles. Scale bar is 20$\mu$m and red (blue) dots indicate positive (negative) $Q$.}
\label{fig:3}
\end{figure}

To simulate colloidal ice we perform Brownian dynamics, carefully parametrized to mimic the experiments~\cite{ortiz-ambriz_engineering_2016}. 
We consider a $2$D array of double wells, each filled by one paramagnetic colloid of diameter $d=10.3\mu m$ and magnetic volume susceptibility ${\chi=0.048}$. 
The overdamped equation of motion for one colloid at position
$\bm{r}_i$ is:
\begin{align}
\gamma \frac{d\bm{r}_i}{dt} = \vec{F}_i^{\mathrm{T}} + \vec{F}_i^{\mathrm{dd}} + \bm{\eta}
\end{align}
where $\gamma = 0.032 \mathrm{pN\ s}\ \mu{}\mathrm{m}^{-1}$, $\vec{F}_i^{\mathrm{T}}$ is the force from the double well which is modelled as a bistable harmonic potential; {\blue more details are in the Supplemental Material~\cite{EPAPS}}. The dipolar force acting on particle $i$ due to the neighbouring colloids is, 
$\vec{F}_i^{\mathrm{dd}} = \frac{3\mu_0}{4\pi} \sum_{j\neq i}  \frac{\vec{m}^2\hat{r}_{ij}}{|r_{ij}|^4}$, where $\vec{m} = \chi V \vec{B}/(\mu_0)$ the dipole moment induced by the external field $\vec{B}$, $\mu_0=4\pi \times 10^{-7} \rm{H/m}$, and $\hat{r}_{ij}=(\vec{r}_i-\vec{r}_j)/|\vec{r}_i-\vec{r}_j|$. {\blue To consider long-range dipolar interactions between the particles, we apply a large cutoff of 200$\mu{}$m.}
Finally, $\bm{\eta}$ represents a random force due to thermal fluctuation, with zero mean, $\left< \bm{\eta} \right> = 0$ and delta correlated, 
$\left< \bm{\eta} \left(t\right) \bm{\eta} \left(t'\right)\right> = 2k_\mathrm{B}T\gamma \delta(t-t')$, with a temperature $T=300\mathrm{K}$. 
Simulations are performed for different system sizes, ranging from $L=3$ to $L=40$, where $L$ is the number of vertices along the side. 
We increase $B$ linearly up to $B= 25$mT, at a rate $0.03125$mT/s.
{\blue For $B= 25$mT, the pair potential between two particles in the closest (farthest) place of two double wells is $U=1176 \, k_B T$ ($U=115 \, k_B T$).}
Our strategy to fix the boundary consists in placing a particle in a single harmonic well at a location 
such that it corresponds to a spin pointing $in$ or $out$.  

We consider four different situations: Two fixed boundary conditions, namely  
antiferromagnetic (AF) and domain wall (DW), illustrated in Fig.~\ref{fig:1}(b). 
In AF boundaries, colloids are placed alternately pointing $in$ and $out$. However, flipping a subset of the colloids in an AF boundary can create defects that are topologically constrained to the inner region, as illustrated in Fig.~\ref{fig:1}(d). 
{\blue This is the basis of the Gauss' law analogue introduced in \cite{king_quantum_2020} for a qbit system. As constructed, the AF state has a neutral charge at the boundaries. This charge neutrality is broken when a spin is changed from \emph{out} to \emph{in} and two defects are created on the AF state.  
With this strategy, we introduce in Fig.~1(d) the Antiferromagnetic Domain Wall (AFDW) where we mix AF boundaries with charged corners.
This configuration produces different behavior with system size $L$. With $L$ is even, two corners point \emph{in}, two point \emph{out}, and the charge is $Q=0$. Instead, with $L$ odd, the four corners point either \emph{in} or \emph{out}, and a total charge $Q=\pm4$ is locked inside the bulk.}
Further we also ran simulations with periodic boundaries, that are similar to previous simulation on particle based ice~\cite{Libal2006,Libal2018}, {\blue and with open boundaries, which represent the experiments with no fixed particles, Fig.~\ref{fig:1}(c).}

To show how borders can be manipulated in experiments, we realize
a square colloidal ice with antiferromagnetic domain walls. The system set-up 
has been described in~\cite{oguz_topology_2020}. Here we modify the boundaries of an isotropic lattice by adding non magnetic silica particles to the corresponding double wells. The silica particles induce local jamming, fixing the paramagnetic particle to a stable location, {\blue as shown below}.

We start by showing in Fig.~\ref{fig:2}(a) how the four different boundary conditions influence the bulk behaviour in terms of the fractions of type III vertices (top) and average vertex charge (bottom).
Both the open and DW frontiers show very similar trends, failing to reach the GS for all sizes.    
For these type of boundaries, the system accumulates charged defects at the boundaries, which are all negative for open boundaries, and positive (negative) for inward-pointing (outward-pointing) spins in the DW.
Only open boundaries allow the appearance of a net non zero topological charge, which converges to a size dependent negative value at high field, as shown in the bottom of Fi g.~\ref{fig:2}(a).
This effect can be appreciated also from the time evolution of the system in Fig.~\ref{fig:2}(b) {\blue and in video S1}.  
Above $B=16.6$mT, all the borders exhibit particles displaced toward the outer region (spins $out$), a radial polarization effect predicted in Ref.~\cite{nisoli_unexpected_2018},
{\blue Such effect arises since the analogy between spin and colloidal ice is broken near the boundaries due to the repulsive interactions between the particles, while in ASIs nanoislands interact due to in-plane dipolar forces.}
In contrast, periodic, AF and DW satisfy the conservation of 
topological charge $Q$ for all field values and system sizes. 
As shown in Fig.~\ref{fig:2}(b), {\blue and videos S2 and S3,}
we find that a system with periodic or DW  boundaries induce the formation of system-spanning domain walls, not allowed by the AF {\blue (video S4)}. 
These defect lines are very difficult to erase by increasing $\bm{B}$ further, as they require the simultaneous flipping of large GS regions in the bulk. 
In a system with periodic boundaries, the parity of the domain walls is also topologically protected: when the boundaries are of even size ($L\in2\mathbb{Z}^{+}$), defect lines can only appear in pairs. In contrast for odd values of $L$, at least one defect line is always present. This effect appears also in Fig.~\ref{fig:2}(c), where the periodic boundaries exhibit a zigzag trend: odd lengths have an excess of type IV vertices which become less relevant as the boundary to bulk ratio becomes smaller.  In contrast we found that AF boundaries can equilibrate to the GS faster and at lower fields since they restrict the phase space as predicted in Ref.~\cite{han_phasespace_2009}. 
{\blue The kinetics of the defects is analyzed in Fig.~\ref{fig:2}(d) for AF boundaries and compared to the periodic case. Both display coarsening dynamics with a power law scaling. This behavior can be also appreciated from the time evolution of the type III domains in Fig.~\ref{fig:2}(e). 
Initially, both systems create similar domain structures, 
but while a system with periodic boundaries falls to a metastable state with several smaller domains, the AF creates a single loop of defects that continuously shrink giving rise to the GS.}

We now explore the behaviour when boundaries are fixed in the AFDW state. Fig.~\ref{fig:3} shows a system with $Q=-4$, where, if the energy of the type IV vertices were similar to that of type III, the charge would be contained in a single type I vertex, with four lines of type IV connecting it to the corners. However, due to line tension, as the applied field increases, it becomes more stable to break up the excess of charges and distribute them along two lines connecting the four corners.
This leads to a symmetry breaking, where the system must choose whether to arrange the two connecting lines horizontally (state $0$ in Fig.~\ref{fig:3}(a,b) {\blue video S5}) or vertically (state $1$, {\blue video S6}). We quantify this bistability using the order parameter 
$\Phi = \left< \left|\vec{s}\cdot{\unit{x}}\right| - \left|\vec{s}\cdot{\unit{y}}\right| \right>$,
where $\vec{s}$ is a sum over the vectors associated to charged vertices, and $\left<...  \right>$ is an average calculated over all vertices. 
By definition, $\Phi$ acquires a positive (negative) value of for defects arranged in the state $0$ (state $1$), Fig.~\ref{fig:3}(b).
As shown in Fig.~\ref{fig:3}(c), we observe a bifurcation starting from $B\sim 5.6$mT.
The process of choosing one of these two states develops via a coarsening 
of small type III domains and consequent reduction of the highly charged defects
until three main domains are formed at $B\sim 9.4$mT. From here, the rest of the process consists of pulling, through line tension, the defect line toward the edges. 

\begin{figure}[t!]
\includegraphics[width = \columnwidth]{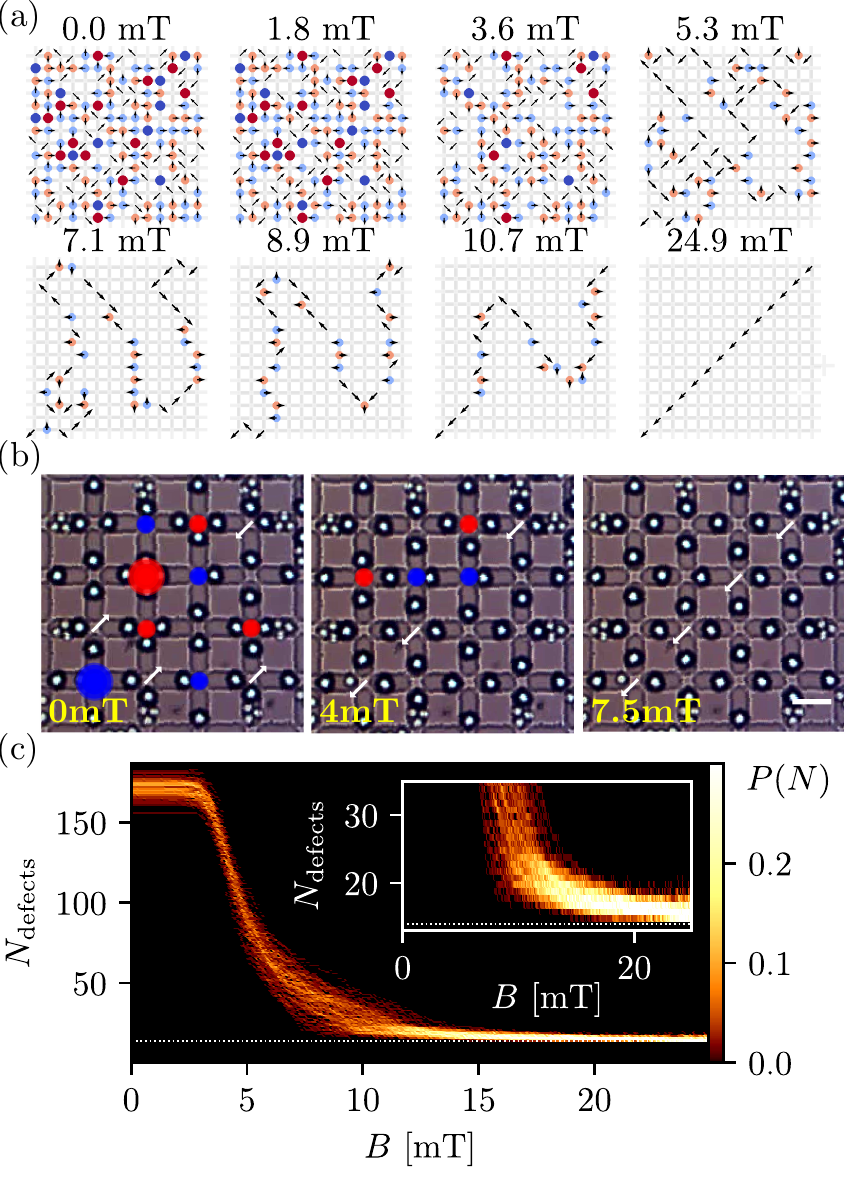}
\caption{(a) Colormap illustrating the time evolution of the topological line connecting two edge defects in the AFDW configuration. (b) Experimental realization of the line of defects using AFDW configuration. Scale bar is 20$\mu$m. (c) The probability of finding a specific number of defects in a system, as a function of field. The dotted line indicates the system's width (L = 14 ). The inset shows a zoom of the central portion of the curve.}
\label{fig:4}
\end{figure}

Another type of AFDW boundary condition can be imposed by introducing only two defects in opposite corners. 
This constraint creates two equal and incompatible type III regions that meet along the diagonal and are separated by a string of type IV vertices. 
The corresponding evolution from a disordered state is shown in Fig.~\ref{fig:4}(a,b) {\blue and in video S7}.
In the simulations, after $B\sim 5$mT, the system nucleates two type III regions which coarsen to the two final domains at $B=25$mT. The final straightening process of the topological string results from line tension, as also confirmed by experiments, Fig.~\ref{fig:4}(c). However, we find that this defect line is not always completely stretched, and defects might appear in the form of small distortions connected by a string of type IV vertices, parallel and pointing along the opposite direction from the main defect line ({\blue Fig.~\ref{fig:1} of the SI \cite{EPAPS}}).
We capture this effect by measuring in Fig.~\ref{fig:4}(c)
the distribution of the number of defects. 
As the field increases, the system gets rid of all non type III vertices, until it reaches a steady-state close to the topologically protected minimum number of defects, which is equal to $L$ (dotted white line in Fig.~\ref{fig:4}(c)). Nevertheless, the inset shows that such minimum value is not reached by many of the realizations, since many systems fall to a state with a small number of defects distributed along the domain wall, and deviating from the diagonal. 

To conclude, we show how to engineer different boundary conditions to control the bulk behaviour
in a geometrically frustrated soft matter system. 
We demonstrate this concept with an artificial colloidal ice combining numerical simulation and experimental realizations.
Topological defects placed at the boundaries propagate inside the bulk, forming 
bistable states with symmetry breaking, or topologically protected strings.
{\blue The fact that the observed phenomena display a topologically protected nature suggest that they could be observed on other systems such as nanomagnetic artificial ice. This could be tested experimentally for example by} using lithography to design smaller and compact islands such as ferromagnetic cubes~\cite{Louis2018}, cylinders~\cite{Sklenar2019}
or disks~\cite{Streubel2018} at the system edges to 
impose a desired bulk configuration.
From the technological perspective, writing or erasing 
defect lines in the GS region can be used to freeze information into the system by applying a bias during the equilibration process~\cite{Wang2018}. 

\acknowledgments{We thank Demian Levis and Leticia Cugliandolo for pointing us towards their work on Arctic regions in vertex models. Experiments were realised with the help of the MicroFabSpace and Microscopy Characterization Facility at IBEC. This project has received funding from the European Research Council (ERC) under the European Union's Horizon 2020 research and innovation programme (grant agreement no. 811234).
P. T. acknowledge support the Generalitat de Catalunya under Program ``ICREA Acad\`emia''.}

\bibliography{bibliography.bib}
\end{document}